\documentclass[prb, twocolumn, showpacs]{revtex4}
\usepackage{amsmath, amssymb, exscale, bm}
\usepackage{array}

\DeclareMathOperator{\const}{const}
\DeclareMathOperator{\im}{Im}
\DeclareMathOperator{\re}{Re}
\newcolumntype{C}{>{$}c<{$}}

\begin{document}

\title{Interference of Instanton Trajectories in Quantum Tunneling for Small
Particles of Real Antiferromagnets}

\author{B. A. Ivanov}
\email{bivanov@i.com.ua}

\author{V. E. Kireev}
\email{kireev@imag.kiev.ua}

\affiliation{Institute of Magnetism NAS of Ukraine, 36-B Vernadskii av., 03142,
  Kiev, Ukraine}

\date{\today}

\begin{abstract}
  For a two-sublattice antiferromagnet the Lagrangian is constructed taking into
  account Berry phase whose form is matched with the quantum-mechanical
  Heisenberg Hamiltonian.  Tunnel effects are analyzed taking into account the
  crystallographic symmetry and possible types of Dzyaloshinskii-Moriya
  interaction.  It is shown that, when the real magnetic symmetry and the
  Dzyaloshinskii-Moriya interaction are taken into consideration, the effects of
  a destructive instanton interference and the suppression of macroscopic
  quantum tunneling can play an essential role.  It also may lead to a periodic
  dependence of the ground-state level splitting on the Dzyaloshinskii-Moriya
  interaction constant; the magnitude of this splitting is calculated.
\end{abstract}


\maketitle


\section{Introduction}

During the last decade, macroscopic quantum tunneling in macroscopic (or, to be
more precise, mesoscopic) magnetic systems has become an object of intense
experimental and theoretical investigations.\cite{ChudnTej98b} In the physics of
magnetism such systems include small magnetic particles, magnetic clusters, and
high-spin molecules.  Special attention is paid to the coherent macroscopic
quantum tunneling (CMQT) between physically different, but energetically
equivalent, states in systems with discrete degeneracy of the ground state.  A
typical CMQT effect in such systems is tunneling between two equivalent
classical states corresponding to two minima of the anisotropy
energy.\cite{QTM95}

The CMQT effects can be observed experimentally from the resonant absorption of
electromagnetic waves at tunnel-splitted energy levels.  The interest in these
effects are associated with the two following factors.  First, mesoscopic
objects exhibiting quantum-mechanical properties are interesting as potential
elements for quantum computers.  Second, fine and elegant effects of
interference of instanton trajectories emerge in these problems.  For
ferromagnetic particles these effects suppress tunneling for half-integral
values of the total spin of the system\cite{LossDiVGrinst92, DelftHenley92} and
lead to oscillatory dependences of the tunnel splitting of energy levels on
extrinsic parameters.\cite{GolPopkov95} In addition, in contrast to the effects
of quantum escape from a metastable to stable state the manifestations of the
CMQT effects are not masked by thermal fluctuations.

Initially, the CMQT investigations were carried out for small particles of a
ferromagnet\cite{Chudn79, ChudnGunter88prb} under the assumption that all spins
in the particle are parallel to one another (the high-spin model).  The effects
of destructive interference of instanton trajectories and interference
suppression of tunneling were predicted precisely for such
systems.\cite{LossDiVGrinst92, DelftHenley92, Aw+92} It turned out later that
antiferromagnets form a more convenient class for experimental investigations of
CMQT.  According to calculations of Refs.\cite{BarbChudn90, KriveZasl90}, the
level spitting in antiferromagnets is stronger than in ferromagnets, and the
effects can be observed at high temperatures.  It is not surprising that the
CMQT effects were observed for the first time in ferritin particles with an
antiferromagnetic structure.\cite{Aw+92} No interference effects are observed in
pure antiferromagnets (i.e., in the case of complete compensation of the spins
of sublattices), but such effects may appear in the applied magnetic
field.\cite{GolPopkov95} It will be shown below that even at zero field the
interference effects can also appear when the real magnetic symmetry of the
crystal is taken into account, in particular, in the presence of the
Dzyaloshinskii-Moriya (DM) interaction.

A semiclassical description of magnetic systems is based on the formalism of
coherent spin states.  In order to construct the effective field Lagrangian both
for ferromagnets and antiferromagnets, we will proceed from the expression for
the Euclidean Lagrangian of an individual spin, which has the
form\cite{Fradkin91}
\begin{equation}\label{lagr:l0}
\mathcal{L}_0 = 
-i \hbar s \sum_k \dot{\phi}_k (1 - \cos\theta_k) + W(\phi_k,\theta_k) \;.
\end{equation}
Here, $s$ is the spin associated with each magnetic moment, $\phi_k$ and
$\theta_k$ are the polar coordinates of the $k$th magnetic moment, and
$W(\phi_k, \theta_k)$ is the classical magnetic energy of the magnet; the
overdot indicates the differentiation with respect to the imaginary time $\tau =
it$.  The first term determines the magnetization dynamics (its variation leads
to the well-known Landau-Lifshitz equations in the angular parametrization) and
also determines a so-called Berry phase.\cite{Fradkin91, Berry84} This quantity
is associated with the total time derivative which is not manifested itself in
the equations of motion, but it is responsible for the interference of instanton
trajectories.

For a macroscopic description it is natural to use one or several field
variables (order parameters) instead of the set of microscopic variables.  The
determination of the number of order parameters and their transformation
properties in magnetic systems is a nontrivial problem.  In the approach based
on coherent spin states the order parameter for ferromagnets is the
magnetization vector of a constant length, which can be parametrized by the
angular variables $\theta$ and $\phi$.  In this case the Berry phase is just a
resultant change in the angle along the instanton
trajectory.\cite{LossDiVGrinst92, DelftHenley92} The behavior of
antiferromagnetic systems can be correctly described using a three-component
vector of a fixed length, viz., the antiferromagnetism vector $\bm{l}$, see
Refs.\cite{AndrMarch80, KosIvKov83b, BarIv92}.  The total spin in this case is a
slave variable, and it is determined by the vector $\bm{l}$ and its time
derivative $\partial\bm{l} / \partial t$.  Dynamic equations for the
antiferromagnetism vector $\bm{l}$ can be either constructed from the symmetry
considerations\cite{AndrMarch80} or derived from the Landau-Lifshitz equations
for the sublattice magnetizations.\cite{BarIv79, Mikeska80} In both these
approaches the same classical equations of motion for the unit vector $\bm{l}$
are obtained, which are usually referred to as the equations of the
$\sigma$-model.  The application of such equations considerably simplifies the
analysis of both linear and nonlinear dynamic effects in an antiferromagnet, see
Refs.\cite{BarIvChetkin85, Bar+94}.  However, the advantage of these equations
for describing macroscopic quantum effects is not so obvious.  The Lagrangian
obtained from the classical Landau-Lifshitz equations or from symmetry
considerations cannot be used directly for describing the MQT effect taking into
account the interference of instanton trajectories.  It is probably for this
reason that Golyshev and Popkov\cite{GolPopkov95} used in their analysis of the
CMQT effects a system of two equations for the sublattice magnetizations, whose
analysis is much more complicated.

As a matter of fact, it is impossible in principle to reconstruct the Lagrangian
of a dynamic system from the classical equations of motion.  The Lagrangians
describing the same classical equations of motion for the system can differ in
the term which is the total derivative with respect to time.  This term does not
effect on the classical dynamics of the system, but alters the magnitude of the
Euclidean action on trajectories.  For this reason the corresponding terms with
total derivatives were lost in the early publications.\cite{Chudn79,
  ChudnGunter88prb} A consistent quantum-mechanical expression for the spin
Lagrangian taking into account the correct equation for the total derivative can
be derived using the formalism of coherent states and the analysis of the
evolution operator; this expression coincides with Eq.~\eqref{lagr:l0} given
above.  Topological terms of the form of total derivatives in the effective
Lagrangian for the vector $\bm{l}$ are significant for the quantum theory of
$1D$ antiferromagnets.\cite{Fradkin91} However, it is impossible in principle to
derive their expressions only from the classical equations of the $\sigma$-model
for the vector $\bm{l}$.

In the simplest version of the $\sigma$-model the derivatives of $\bm{l}$ with
respect to time appear in the Lagrangian in the trivial form $(\partial\bm{l} /
\partial\tau)^2$, see Refs.\cite{AndrMarch80, KosIvKov83b, BarIv92}.  In this
case the equations of the $\sigma$-model are Lorentz-invariant, and the
description of the dynamics of nonlinear magnetization waves (kink-type solitons
in antiferromagnets) is considerably simplified.\cite{BarIv92, BarIvChetkin85}
The interference effects in the MQT are obviously absent.  However, the
situation changes drastically for more realistic models.

First, for many antiferromagnetic crystals there exist terms reflecting
interactions of the DM type, which are linear in $\bm{l}$ and in magnetization.
It was shown in Refs.\cite{Hom+89, Hom+90} that these interactions are
responsible for the terms in the effective Lagrangian which are linear in
$\partial\bm{l} / \partial\tau$; this considerably modifies the kink dynamics in
comparison with the simplest Lorentz-invariant model.  Obviously, such
interactions can also lead to the emergence of total derivatives (topological
phases).  The presence of a magnetic field may also lead to similar effects;
this was noted in the analysis of the nonlinear dynamics of
antiferromagnets\cite{IvKolOksyuk91} as well as for the MQT effect, see
Ref.\cite{IvKolOksyuk91} and recent publications\cite{GolPopkov95, IvKir99,
  Lu+98, Lu+00-1, Lu+00-2, Lu+00-3}.

In the present paper we will construct the Lagrangian of the $\sigma$-model on
the base of the Eq.~\eqref{lagr:l0} taking into account all possible sources of
the terms with the total derivative, which may lead to nontrivial interference
effects.  This Lagrangian will be used to study the interference of instanton
trajectories for the real models of antiferromagnetic particles of various
symmetries and to determine the contribution of these effects to the tunneling
probability.


\section{Lagrangian of the $\bm{\sigma}$-model for real antiferromagnets}
\label{s:lagr}

Let us consider a system with localized spins, in which nearest neighbors are
coupled through the antiferromagnetic interaction.  We assume that the lattice
has such a structure that the sites with spins can be divided into two groups so
that the spins appearing in pairs of nearest neighbors belong to different
groups and there are no frustrations in the lattice.  For ideal antiferromagnets
these two groups correspond to the two magnetic sublattices.  In this case we
assume that the spins corresponding to each group have parallel orientations and
form the total spins $\bm{S}_1$ and $\bm{S}_2$ of the sublattices.  In the
exchange approximation for such antiferromagnets, vectors $\bm{S}_1$ and
$\bm{S}_2$ are antiparallel.  The total spin $\bm{S} = \bm{S}_1 + \bm{S}_2$ in
the ground state can differ from zero in view of a different number of sites in
the sublattices (decompensation), $|\bm{S}_1| \neq |\bm{S}_2|$, and also in the
presence of an external magnetic field and/or the DM interaction, when the
antiparallelism of the spins is violated (i.e., $|\bm{S}_1 + \bm{S}_2| \neq 0$
even for $|\bm{S}_1| = |\bm{S}_2|$). We will consider only completely
compensated antiferromagnets with $|\bm{S}_1| = |\bm{S}_2|$, since the specific
effects associated with spin decompensation ($|\bm{S}_1| \neq |\bm{S}_2|$, but
the difference $|\bm{S}_1 - \bm{S}_2| \ll |\bm{S}_{1,2}|$) reduce the
interference effects to those which are well known for
ferromagnets.\cite{ChiolLoss97}

Our goal is to construct the Lagrangian describing the dynamics of the vector
$\bm{l}$ in the presence of the DM interaction and the magnetic field.  Since
the tensor of exchange interaction constants $J_{i j}$ may have the
antisymmetric component in the nearest neighbor approximation, the Hamiltonian
of such a system has the form
\begin{equation}\label{lagr:exch}
\mathcal{H}_{e} = 
J \sum_{<\alpha \beta>} \bm{S}_\alpha \bm{S}_\beta + 
\sum_{<\alpha \beta>} \bm{d} \cdot (\bm{S}_\alpha \times \bm{S}_\beta) - 
g \mu_B \sum_\alpha \bm{H} \cdot \bm{S}_\alpha \;.
\end{equation}
Here, the first term describes the isotropic exchange interaction, the summation
in this term is extended to the pairs of nearest neighbors, and $\bm{S}_\alpha$
is the spin at the $\alpha$th site.  The antisymmetric component of the tensor
of the exchange constants $J_{i j}$ is a microscopic source of the DM
interaction,\cite{Moriya60} corresponding to the dual vector $\bm{d}$.  The last
term describes the interaction of spins with the external magnetic field.

Let us consider the exchange approximation in which the deviation from the
conventional Heisenberg model with an isotropic exchange interaction $J
\bm{S}_\alpha \bm{S}_\beta$ is small; i.e., $d$, $g \mu_B H \ll J$.  In this
case we can introduce the total spins $\bm{S}_1 = \sum \bm{S}_{\alpha_1}$ and
$\bm{S}_2 = \sum \bm{S}_{\alpha_2}$ of the sublattices and assume that the
vectors $\bm{S}_1$ and $\bm{S}_2$ have a fixed length.  It is convenient to put
$\bm{S}_1 = N s \bm{\sigma}_1$ and $\bm{S}_2 = N s \bm{\sigma}_2$, where $s$ is
the spin at a site and $N$ is the number of sites in each sublattice.  We will
parametrize the unit vectors $\bm{\sigma}_1$ and $\bm{\sigma}_2$ by the polar
coordinates $(\theta_1, \phi_1)$ and $(\theta_2, \phi_2)$, respectively.  In
this case the classical magnetic energy of the antiferromagnet, whose exchange
component corresponds to the Hamiltonian~\eqref{lagr:exch}, can be written in
the form
\begin{multline}\label{lagr:lagr1}
\mathcal{W}(\bm{\sigma}_1, \bm{\sigma}_2) = 
J s^2 z N \bm{\sigma}_1 \bm{\sigma}_2 +
s^2 z N \bm{d} \cdot (\bm{\sigma}_1 \times \bm{\sigma}_2) \\
+ w(\bm{\sigma}_1, \bm{\sigma}_2) 
- g \mu_B s N \bm{H}(\bm{\sigma}_1 + \bm{\sigma}_2) \;.
\end{multline}
Here, $N$ is the number of spins in sublattices, $z$ is the coordination number
for a lattice site, and $w(\bm{\sigma}_1, \bm{\sigma}_2)$ is the anisotropy
energy.

Thus, we arrive at the description of the energy of an antiferromagnet in terms
of two unit vectors.  Their dynamics can be described by a Lagrangian which can
be written in the dynamic variables $\bm{\sigma}_1$ and $\bm{\sigma}_2$ taking
into account relation~\eqref{lagr:lagr1} in the form
\begin{equation}\label{lagr:l-v}
\mathcal{L} = 
-i \hbar S_1 \bm{A}_1 (\bm{\sigma}_1) \dot {\bm{\sigma}_1} - i\hbar
S_2 \bm{A}_2 (\bm{\sigma}_2) \dot{\bm{\sigma}_2} -
\mathcal{W}(\bm{\sigma}_1, \bm{\sigma}_2) \;.
\end{equation}
Here, we have chosen a more general form of the kinetic terms as compared to
relation~\eqref{lagr:l0}.  These terms can be presented through the vector
potential of the field of a magnetic monopole:
\begin{equation}\label{lagr:mono}
\bm{A}_{1, 2}(\bm{\sigma}) = 
\frac{\bm{\sigma} \times \bm{n}_{1, 2}}
{\sigma(\sigma + \bm{\sigma} \bm{n}_{1, 2})} \;,
\end{equation}
where $\bm{n}_{1, 2}$ are the quantization axes of coherent states for each
sublattice.  This potential has a singularity for $\bm{\sigma}\bm{n} = -\sigma$,
i.e., on a certain half-line in the $\bm{\sigma}$-space.  Usually, the ``north
pole'' gauge with $\bm{n} = \hat{\bm{e}}_z$ is used, in which the quantity
$\bm{A}(\bm{\sigma}) \dot{\bm{\sigma}}$ assumes to have the familiar form
\eqref{lagr:l0}.  The potentials $\bm{A}_{1, 2}$ of the monopole field permit
gauge transformations (such as a change in the position of spin quantization
axes and, hence, singularities), which do not change the equations of motion,
but make a contribution to the Lagrangian in the form of the total derivative of
the function of spins $\bm{\sigma}_1$ and $\bm{\sigma}_2$ with respect to
$\tau$, which can be significant for the description of interference effects.
The kinetic terms for each sublattice can be written in individual gauges (in
particular, with different directions of the quantization axes $\bm{n}_1$ and
$\bm{n}_2$).

We present the unit vectors $\bm{\sigma}_1$ and $\bm{\sigma}_2$ in terms of
vectors $\bm{l} = (\bm{\sigma}_1 - \bm{\sigma}_2) / 2$ and $\bm{m} =
(\bm{\sigma}_1 + \bm{\sigma}_2) / 2$ which are connected through the relations
\begin{equation}
\bm{l}^2 + \bm{m}^2 = 1 \qquad \text{and} \qquad \bm{ml} = 0 \;.
\end{equation}

In the $\sigma$-model approximation, when the magnetic moment is small $|\bm{m}|
\ll |\bm{l}|$, we can use simple transformations in order to present the
Lagrangian \eqref{lagr:l-v} as a power expansion in $\bm{m}$.  Confining the
expansion to the terms linear in $\bm{m}$, we can write the kinetic terms in the
form
\begin{multline}\label{lagr:kin}
-i \hbar \bm{A}_1 (\bm{\sigma}_1) \dot {\bm{\sigma}_1} -
i \hbar \bm{A}_2 (\bm{\sigma}_2) \dot {\bm{\sigma}_2} = \\
 = -i \hbar \dot{\bm{l}} [\bm{A}_1(\bm{l}) - \bm{A}_2(-\bm{l})] -
i \hbar m_i \biggr[\dot{\bm{l}} \frac{\partial\bm{A}_1(\bm{l})} {\partial l_i} +
\dot{\bm{l}} \frac{\partial\bm{A}_2(-\bm{l})} {\partial l_i}\biggl] \\
-i \hbar \dot{\bm{m}} [\bm{A}_1(\bm{l}) + \bm{A}_2(-\bm{l})] \;.
\end{multline}
Here, the form of the vector-potentials $\bm{A}_1$ and $\bm{A}_2$ has not been
specified yet.  In particular, the quantization axes $\bm{n}_1$ ans $\bm{n}_2$
have not been chosen.  It is natural to choose the quantization axes $\bm{n}_1$
and $\bm{n}_2$ so that $\bm{A}_1(\bm{l}) = \bm{A}_2(-\bm{l})$, which is possible
for $\bm{n}_1 = -\bm{n}_2$.  In this case the singular term with $d\bm{l} / d
\tau$ vanishes, and the dynamic terms expansion starts with the term linear in
$\bm{m}$, which can be written in the form
\begin{equation}\label{lagr:kin2}
-i \hbar \bm{m} \cdot (\bm{F} \times \dot{\bm{l}})
-i \hbar \frac{d}{d\tau}[\bm{m} \cdot \bm{A}_1(\bm{l})] \;,
\end{equation}
where
\begin{equation}
F_i = 
\epsilon_{i j k} \biggl(\frac{\partial A_j} {\partial l_k} - 
\frac{\partial A_k} {\partial l_j}\biggr) \;.
\end{equation}
Thus, most of the arbitrariness in the choice of the gauge field $\bm{A}$, which
takes place for ferromagnets, does not exist for antiferromagnets.  The
gauge-invariant quantity $F_i$, which has the meaning of a formal magnetic field
associated with potential $\bm{A}$, is the magnetic monopole field $\bm{F} =
\bm{l} / |\bm{l}|^3$.  In the transition from Eq.~\eqref{lagr:kin} to the
expression~\eqref{lagr:kin2} the initial gauge arbitrariness turned out to be
localized in the term with the total derivative $d [\bm{m} \bm{A}(\bm{l})] / d
\tau$.  Concerning this quantity, its contribution to the Euclidean action is
obviously equal to zero in the case when an instanton trajectory misses the
singular point of $\bm{A}(\bm{l})$.  This condition can be easily satisfied if
we choose the direction $\bm{n} = \bm{n}_1 = -\bm{n}_2$ along the hard
magnetization axis of the antiferromagnet.  In this case the phase for a closed
path on the unit sphere $\bm{l}^2 = 1$, which is formed by instanton
trajectories, is independent of the position of the quantization axis $\bm{n}$.

Taking into account the condition $\bm{ml} = 0$, we eliminate from the
expression~\eqref{lagr:l-v} the slave variable $\bm{m}$:
\begin{equation}\label{lagr:m}
\bm{m} = 
\frac{\hbar} {2 J s z} \biggl[\gamma (\bm{H}^\text{eff} - \bm{l}
(\bm{H}^{\rm eff} \cdot \bm{l})) - i\bm{l} \times \dot{\bm{l}}\biggr] \;,
\end{equation}
where $\gamma = g \mu_B / \hbar$ is the gyromagnetic ratio and
$\bm{H}^\text{eff}$ is the effective field which is the sum of the external
field $\bm{H}$ and the DM field $\bm{H}^\text{DM}$.  In the approximation chosen
above, in which the DM interaction can be presented in the purely antisymmetric
form $\bm{d} (\bm{\sigma}_1 \times \bm{\sigma}_2) \propto \bm{d}(\bm{l} \times
\bm{m})$, the DM field can be written as $\bm{H}^\text{DM} = z s (\bm{d} \times
\bm{l}) / (g \mu_B)$.

The expression for $\bm{m}$ is also valid for more general forms of the DM
interaction, which cannot be reduced to a bilinear form in $\bm{\sigma}_{1, 2}$.
In particular, we will consider more general forms of the DM interaction of the
type $D_{i k} (\bm{l}) m_i l_k$ which are observed for many crystals and are
significant for the MQT effects.  In this case the effective field in the
expression for $\bm{m}$ assumes the form $H^\text{DM}_i = D_{i k}(\bm{l}) l_k$,
and
\begin{equation}\label{lagr:Hd}
H^\text{eff}_i = 
H^{(0)}_i + D_{i k}(\bm{l})l_k \;.
\end{equation}

In this section we will not specify the form $D_{i k}(\bm{l})$.  The
approximation $|\bm{m}| \ll |\bm{l}|$ used in the derivation of the
$\sigma$-model is satisfied for $\max(H, H^\text{DM}) \ll H_{ex}$, where $H_{ex} = J
s z / \mu_B$ is the exchange field.  Substituting $\bm{m}$ into the
Lagrangian~\eqref{lagr:l-v}, we obtain the effective Lagrangian for the vector
$\bm{l}$ in the form
\begin{multline}\label{lagr:l-eff}
\mathcal{L} = 
\frac{\hbar^2 N}{2J z} \biggl[\frac{1}{2} \dot {\bm{l}}^2 +
i \gamma \bm{H}^\text{eff} \cdot (\bm{l} \times \dot{\bm{l}})\biggr] -
\mathcal{W}_a(\bm{l}) \\
+ \frac{2 \mu_B^2 N}{J z} \biggl\{(\bm{H} \cdot \bm{l})^2 - \bm{H}^2 +
2 \bm{H} \biggl[\bm{l} (\bm{H}^\text{DM} \cdot \bm{l}) - \bm{H}^\text{DM}
\biggr]\biggr\} \;.
\end{multline}
Here, $\mathcal{W}_a(\bm{l})$ has the meaning of the effective anisotropy energy
in which the additional term is $(\bm{H}^\text{DM} \cdot \bm{l})^2 - (\bm{H}^\text{DM})^2$
taken into account along with the initial energy $w(\bm{l}) = w(\bm{\sigma}_1,
\bm{\sigma}_2)$ introduced above for $\bm{\sigma}_1 = -\bm{\sigma}_2 = \bm{l}$.
The quantity $\mathcal{W}_a(\bm{l})$ is obviously the real anisotropy energy
determined from static measurements in weak fields, and there is no point in
separating these contributions.  We must simply use the expression for
$\mathcal{W}_a(\bm{l})$ which is determined by the crystal symmetry of the
magnet.  The specific form of the anisotropy energy for various antiferromagnets
is given in the Table~\ref{t:dz-an}.  The terms in the braces describe the
variation of the static energy of the antiferromagnet due to the external
magnetic field.  The first term, which is quadratic in the components of
$\bm{H}$, is quadratic in $\bm{l}$ and can also be presented as the field
induced renormalization of the anisotropy energy.  The second term, which is
bilinear in the components of the external magnetic field $\bm{H}$ and the DM
field $\bm{H}^\text{DM}$, contains odd powers of the components of $\bm{l}$ and
describes the energy of the weak ferromagnetic moment induced by the DM
interaction.  (In particular, this term can be reduced to $\bm{H}\cdot (\bm{d}
\times \bm{l})$ for a purely antisymmetric DM interaction.) This term can
completely remove the degeneracy of the classical ground state of the system,
and the analysis of the MQT effects becomes meaningless.  For this reason, it
makes sense to take into account the external magnetic field and the DM
interaction simultaneously only for certain selected orientations of the
external field, when this term vanishes for vector $\bm{l}$ directed along the
easy magnetization axis of the antiferromagnet.  Some of these orientations of
the field for orthorhombic antiferromagnets were considered in
Ref.\cite{GolPopkov95}.

Thus, we arrive at the following conclusions.  The Lagrangian for the vector
$\bm{l}$ differs from the Lagrangian for the $\sigma$-model of ideal
antiferromagnets\cite{Fradkin91} in the presence of a number of additional terms
which play different roles in the description of CMQT.  In contrast to the case
of ferromagnets or antiferromagnets with different spins of the sublattices, the
term with the total derivative can easily be eliminated.  It is important to
note that the external field and some forms of the DM interaction leads to the
emergence of gyroscopic terms linear in $d \bm{l} / d \tau$.  The emergence of
these terms indicates the lowering of the actual dynamic symmetry of
antiferromagnets in the presence of a magnetic field and/or the DM interaction.

The structure of the Lagrangian is such that the contribution of the DM
interaction to the gyroscopic term can be taken into account by adding the DM
field $\bm{H}^\text{DM}$, which is a function of $\bm{l}$, to the external
magnetic field $\bm{H}$.  Gyroscopic terms can make significant contributions to
the probabilities of tunneling processes both by effecting on the structure of
instanton solutions and by creating destructive interference of instanton
trajectories.  It will be proved below that, in contrast to the case of a
ferromagnets or antiferromagnets with different spins of the sublattices, this
interference is not of topological origin, but can also be given below. The
examples of ``pure'' antiferromagnets in which tunneling can be completely
suppressed due to the interference of instanton trajectories will be given
below.

Deriving the Lagrangian~\eqref{lagr:l-eff}, we neglected the possibility of
inhomogeneous tunneling and, hence, the dependence of $\bm{S}$ and $\bm{l}$ on
spatial coordinates was omitted from the very outset.  The inclusion of such a
dependence leads to the substitution $N \to \int d V / a^3$, where $a$ is the
lattice constant, and to the emergence of an additional term proportional to $J
a^2 (\nabla\bm{l})^2$ in the Euclidean action.  A comparison of the
inhomogeneity energy with the anisotropy energy leads to an estimate of the
spatial inhomogeneity size on the order of $\Delta_0 = a \sqrt{H_{ex} /
  H_{an}}$, where $H_{ex}$ and $H_{an}$ are the exchange field and the
anisotropy field and $\Delta_0$ is the domain wall thickness.  If the size of a
particle is larger than $\Delta_0$, i.e., $N > N_c \simeq (\Delta_0 / a)^3
\simeq (H_{ex} / H_{an})^{3/2}$, we can assume a more advantageous inhomogeneous
tunneling scenario, in which the level splitting weakly depends on $N$ (or is
even independent of it) for $N > N_c$ Although this question has not been
discussed in the literature and its analysis is beyond the scope of the present
publication, we will briefly consider it.

The value of $N_c$ is too large for the tunneling effects to be observable for
$N > N_c$.  As a matter of fact, $N$ in the tunneling exponent is multiplied by
the susceptibility of the system, i.e., appears in the combination $N H_{an} /
H_{ex}$, see Ref.\cite{KriveZasl90}.  The presence of this small parameter
actually makes it possible to observe tunneling at ferritin particles with $N
\simeq 3.5 \cdot 10^3$, see Refs.\cite{Aw+92, Tej+97}.  However, for typical
values of $H_{an} / H_{ex} \sim 10^{-2}$ -- $10^{-3}$, the tunneling exponent
$N_c H_{an} / H_{ex} \simeq (H_{ex} / H_{an})^{1/2} \gg 1$ is too large and the
observation of the transition to the inhomogeneous tunneling mode becomes
problematic.


\section{Symmetry of the instanton solutions and interference of contributions
from instanton trajectories}\label{s:berry}

In accordance with the general rules of the semiclassical approximation
formulated in the instanton language, the amplitude of transition from one
state to another is described in the so-called instanton-gas
approximation\cite{Vaj+82}.  The level splitting for a system with two
equivalent minima can be presented in the form
\begin{equation}\label{berry:split}
\Delta \propto
D \sqrt{K} \;,
\end{equation}
where the quantity $D$ is defined as
\begin{equation}\label{berry:D}
D = 
(\det{'}\Hat{\Omega})^{-1/2} 
\biggl(\frac{\re\mathcal{I}} {2 \pi \hbar}\biggr)^{1/2}
\exp \biggl(-\frac{\re\mathcal{I}}{\hbar} \biggr) \;,
\end{equation}
and $\mathcal{I}$ is the one-instanton action; $K$ is a combinatorial factor
emerging due to nonuniqueness of the tunnel path connecting two equivalent
minima; and $\det{'}\Hat{\Omega}$ is the fluctuation determinant disregarding
the zeroth mode, which is determined by small deviations from an instanton
trajectory, see Ref.\cite{Vaj+82} for details.  In order to analyze the effects
of tunneling between degenerate states corresponding to the ground states of the
system and to determine the value of splitting, we must find the one-instanton
trajectories connecting these states, calculate the value of the Euclidean
action $\mathcal{I}$ on these trajectories, and find the determinant of the
operator for the second variation of action.  The contribution to the splitting
comes only from equivalent trajectories corresponding to the minimum value of
the real component of $\mathcal{I}$.  The combinatorial factor depending on the
phase difference in the trajectories will be calculated below using
Eq.~\eqref{berry:comb}.  In this section, we concentrate our attention on an
analysis of the main contribution which comes only from $\mathcal{I}$ and will
not calculate the preexponential factor.  Let us see how these calculations can
be carried out in actual practice.

For a concrete analysis, it is convenient to write the Lagrangian in the form
\begin{equation}\label{berry:lagr}
\mathcal{L} = 
\frac{\hbar^2 N}{2 J z} \biggl[\frac{1}{2} \biggl(\frac{d \bm{l}}{d\tau}
\biggr)^2 +
i(\bm{\omega}_H \times \bm{l}) \cdot \frac{d\bm{l}}{d\tau} + 
\frac{\omega_0^2}{2} w_a(\bm{l})\biggr] \;,
\end{equation}
where $\bm{\omega}_H = \gamma \bm{H}^\text{eff}$, $\gamma$ is the gyromagnetic
ratio, and $\bm{H}^\text{eff}$ is the effective field.  The dimensionless
function $w_a(\bm{l})$ is proportional to the anisotropy energy, and the value
of $\omega_0$ coincides with the frequency of a homogeneous antiferromagnetic
resonance in the uniaxial anisotropy field.  We parametrize the vector $\bm{l}$
by the angular variables
\begin{equation}\label{berry:l}
l_1 = \sin\theta \cos\phi \;, \qquad 
l_2 = \sin\theta \sin\phi \;, \qquad
l_3 = \cos\theta \;.
\end{equation}
We are dealing with an easy-axis anisotropy. Consequently, the ground state is
doubly degenerated and has two values of $\bm{l}$ corresponding to it: $\bm{l} =
\hat{\bm{e}}_3$ and $\bm{l} = -\hat{\bm{e}}_3$, and the unit vector
$\hat{\bm{e}}_3$ being parallel to the easy axis.  Let us consider the tunneling
between these two states.  Function $w_a(\bm{l})$ for a magnet with the
anisotropy axis $C_n$ can be written in the form
\begin{equation}\label{berry:anizo}
w_a(\theta, \phi) = 
\sin^2\theta + \widetilde{w}_a(\theta, \phi) \;,
\end{equation}
where the first term corresponds to easy-axis anisotropy and
$\Tilde{w}_a(\theta, \phi) \ll 1$ defines anisotropy in the basal plane.

For antiferromagnets with an easy axis of symmetry $C_n$, there exist $n$
instanton trajectories and $n$ antiinstanton trajectories, and the combinatorial
factor has the form
\begin{equation}\label{berry:comb}
K = 
\sum_{k,\bar{k}' = 0}^{n - 1} \cos\Phi_{k, \bar{k}'} \;, \qquad
\Phi_{k,\bar{k}'} = 
\frac{1}{\hbar} \im \oint_{k\bigcup \bar{k}'} d\tau\, 
\mathcal{L}(\bm{l}, \dot{\bm{l}}) \;,
\end{equation}
i.e., $\Phi_{k, \bar{k}'}$ is the phase difference between the $k$th instanton
and the $k'$th antiinstanton.  The integral defining $\Phi_{k, \bar{k}'}$ is
taken over a closed path formed by the trajectories of the $k$th instanton and
the $\bar{k}'$th antiinstanton. In the Lorentz-invariant $\sigma$-model, the
Lagrangian is real and all $ \Phi_{k, \bar{k}'}$ are equal to zero; the
combinatorial factor $K = n^2$ is trivial and equal to $n$.  Consequently,
$\sqrt{K} = n$; i.e., the total transition amplitude and level splitting for
$n$n pairs is just the contribution from one instanton multiplied by the number
of paths.  It will be shown below, however, that for $\Phi_{k, \bar{k}'} \neq
0$, the level splitting $\Delta$ may contain an oscillatory dependence on the
product of the small parameter $|\bm{\omega}_H|$ and the large quantity $N$,
and, hence, requires a more detailed analysis.  The nature of its oscillations
can be established from symmetry considerations, and the specific form of the
function $K$ of the parameters of the problem can be determined even without
solving the corresponding the Euler-Lagrange equations.

\subsection{Lorentz-invariant $\sigma$-model}

It is convenient to consider first the tunneling in the simplest
Lorentz-invariant $\sigma$-model which corresponds to the
Lagrangian~\eqref{berry:lagr} with $\bm{\omega}_H = 0$.  As a matter of fact,
for some models of an antiferromagnet with the DM interaction, the results turn
out to be the same as in the absence of this interaction, see below.  If $H^{\rm
  eff} = 0$, the analysis of the problem does not present any difficulty.
Indeed, for any form of the anisotropy energy in a uniaxial antiferromagnet with
the principal axis $C_2$, $C_4$, $C_6$ (in the subsequent analysis, we will
consider only the type of symmetry that can exist in the crystal lattice), the
instanton solution corresponds to the function $\theta = \theta(\tau)$ with the
boundary conditions $\theta \to 0,\pi$ for $\tau \to \pm \infty$ and $\phi =
\phi_0 = \const$, where $\phi_0$ is defined by the relation
\begin{equation}
\frac{\partial w_a(\theta, \phi)}{\partial\phi}\biggm|_{\phi = \phi_0} = 0 \;.
\end{equation}

Let us assume that the ground states $\pm \hat{\bm{e}}_3$ are on the principal
axis $C_n$.  In this case the value of $\widetilde{w_a}(\theta, \phi)$ is
proportional to $\sin n\phi$ and there exist $2 n$ solutions to this equation:
\begin{equation}
\phi_k^{(0)} = 
\frac{\pi k}{n} \;, \qquad k = 0, 1, \ldots, 2n - 1 \;,
\end{equation}
from which $n$ solutions $\phi_{k, \text{min}}^{(0)}$ correspond to the minima
of $w_a(\theta, \phi)$, while the remaining $n$ solutions $\phi_{k,
  \text{max}}^{(0)}$ correspond to the maxima of this function for all $\theta
\neq 0, \pi$.  Instantons with $\phi_{k, \text{min}}^{(0)}$ correspond to the
lowest value of the Euclidean action, and we will consider below only these $n$
solutions.  Function $\theta(\tau)$ can be determined from the second-order
equation for which the first integral is known to be
\begin{equation}\label{berry:eul-li}
\biggl(\frac{d\theta}{d\tau} \biggr)^2 = 
\omega_0^2[w_a(\theta, \phi_k^{(0)}) - w_a(0, \phi_k^{(0)})] \;.
\end{equation}

Henceforth, we assume that $w_a(0,\phi) = w_a(\pi,\phi) = 0$ and that the value
of $\phi = 0$ corresponds to the minimum of the function $w_a(\theta, \phi)$.
With such a choice of the axes, $z$ is always an easy magnetization axis and $x$
is an medium magnetization axis.  The Euclidean action on trajectories is
real-valued for all values of $\phi$ and is defined as
\begin{equation}
\mathcal{I} = 
\frac{\hbar^2 \omega_0 N}{2 J z} \int_0^\pi d\theta \, \sqrt{w_a(\theta, \phi)} \;.
\end{equation}

This approximate expression is written in the main approximation in small
anisotropy in the basal plane $\widetilde{\beta} \ll 1$, where
$\widetilde{\beta}$ is the characteristic anisotropy constant in the basal
plane, i.e., the maximum value of $\widetilde{w_a}$ .  Thus, the contribution in
the given case comes from $n$ instanton trajectories on which vector $\bm{l}$ is
real and rotates in one of the $n$ planes defined by the condition $\phi =
\phi_{k, \text{min}}^{(0)} \equiv \phi_k^{(0)}$.  The imaginary component of
$\mathcal{I}$ in the Lorentz-invariant model is absent, and the combinatorial
factor $K$ in the expression~\eqref{berry:comb} is equal to $n^2$.

\subsection{Role of $\bm{H}^\text{eff}$}

The inclusion of the terms with $\bm{H}^\text{eff}$, which destroy the Lorentz
invariance, brings about two types of difficulties.  First, for $\bm{H}^{\rm
  eff} \neq 0$, the solution $\phi = \const$ is generally inapplicable, and the
instanton structure is determined by the general system of two second-order
equations
\begin{subequations}\label{berry:eul}
\begin{equation}\label{berry:eul-th}
-\ddot{\theta} + \dot{\phi}^2 \sin\theta \cos\theta +
\omega_0^2 \frac{\partial w_a}{\partial\theta} +
i \omega_H \dot{\phi} \Gamma(\theta, \phi) = 0 \;,
\end{equation}
\begin{equation}\label{berry:eul-phi}
-\ddot{\phi} \sin^2\theta - 2 \dot{\phi} \dot{\theta} \sin\theta \cos\theta +
\omega_0^2 \frac{\partial w_a}{\partial\phi} -
i \omega_H \dot{\theta} \Gamma(\theta, \phi) = 0 \;,
\end{equation}
\end{subequations}
whose solutions are generally not real-valued.  Here, the terms with $\Gamma$
are determined by the variation of the term with $\bm{H}^\text{eff} \cdot
[(d\bm{l} / d\tau) \times \bm{l}]$ in the Lagrangian~\eqref{berry:lagr}, and the
form of the function $\Gamma(\theta, \phi)$ generated by the DM interaction for
various types of magnetic symmetry is given in the column 5 of the
Table~\ref{t:dz-an}.  Second, the imaginary component of the Euclidean action
$\mathcal{I}$, which comes from the term proportional to $\bm{\omega}_H$, may
appear even for trajectories with a real $\bm{l}$.  Let us consider the cases
when these situations are realized.

\renewcommand{\arraystretch}{2}
\setlength{\extrarowheight}{-2pt}

\begin{table*}
\caption{Anisotropy in the basal plane and the Dzyaloshinskii-Moriya interaction
  constant for systems with various types of the magnetic symmetry.\label{t:dz-an}}
\begin{ruledtabular}
\begin{tabular}{|C|C|C|C|C|C|C|C|}
1 &
2 &
3 &
4\footnote{For high-order axes, the following notation is introduced:
$m_\pm = m_x \pm i m_y$ and $l_\pm = l_x \pm i l_y$.} &
5 &
6\footnote{Asterisks mark systems for which an exact solution
corresponding to the munimum of the real part of the action exist.} &
7 &
8 \\
\hline

n &
\widetilde{w}_a  &
\text{Axes} &
\text{DMI}  &
\Gamma(\theta,\phi) &
&
B(\theta,\phi) &
K    \\
\hline\hline

&
&
2^{(+)}_z 2^{(-)}_x 2^{(-)}_y &
m_x l_y + m_y l_x &
3 \sin^3\theta \sin2\phi  &
* &
3 \sin^2\theta \sin2\phi  &
4    \\
\cline{3-8}

2 &
\beta_2 \sin^2\theta \sin^2\phi &
2^{(-)}_z 2^{(+)}_x 2^{(-)}_y &
m_y l_z + m_z l_y &
6 \sin^2\theta \cos\theta \sin\phi &
* &
6 \sin\theta \cos\theta \sin\phi &
4    \\
\cline{3-8}

&
&
2^{(-)}_z 2^{(-)}_x 2^{(+)}_y &
m_x l_z + m_z l_x &
6 \sin^2\theta \cos\theta \cos\phi &
&
6 \sin\theta \cos\theta \cos\phi &
4    \\
\hline\hline

&
&
4^{(+)}_z 2^{(-)}_x 2^{(-)}_{xy} &
\frac{1}{2i} (m_+ l_+^3 - m_- l_-^3) &
5 \sin^5\theta \sin4\phi &
* &
5 \sin^4\theta \sin4\phi &
16     \\
\cline{3-8}

4 &
\beta_4 \sin^4\theta \sin^2 2\phi &
4^{(-)}_z 2^{(+)}_x 2^{(-)}_{xy} &
m_x l_x - m_y l_y &
3 \sin^3\theta \cos 2\phi &
&
3 \sin^2\theta \cos 2\phi &
16   \\
\cline{3-8}

&
&
4^{(-)}_z 2^{(-)}_x 2^{(+)}_{xy} &
m_x l_y + m_y l_x &
3 \sin^3\theta \sin 2\phi &
* &
3 \sin^2\theta \sin 2\phi &
8 + 8 \cos(s d N / J)  \\
\hline\hline

&
&
6^{(+)}_z 2^{(-)}_x 2^{(-)}_{\pi/6} &
\frac{1}{2i}(m_+ l_+^5 - m_- l_-^5) &
7 \sin^7\theta \sin 6\phi &
* &
7 \sin^6\theta \sin 6\phi &
36   \\
\cline{3-8}

6 &
\beta_6 \sin^6\theta \sin^2 3\phi &
6^{(-)}_z 2^{(+)}_x 2^{(-)}_{\pi/6} &
\frac{1}{2i} m_z (l_+^3 - l_-^3) &
5 \sin^4\theta \cos\theta \sin3\phi &
* &
5 \sin^3\theta \cos\theta \sin3\phi &
36  \\
\cline{3-8}

&
&
6^{(-)}_z 2^{(-)}_x 2^{(+)}_{\pi/6} &
\frac{1}{2} m_z (l_+^3 + l_-^3) &
5 \sin^4\theta \cos\theta \cos3\phi &
&
5 \sin^3\theta \cos\theta \cos3\phi &
36  \\

\end{tabular}
\end{ruledtabular}
\end{table*}

If $\Gamma(\theta, \phi)$ vanishes at the same values of $\phi_k^{(0)}$ as for
$\partial w_a(\theta, \phi)/\partial\phi$, the second
equation~\eqref{berry:eul-phi} is satisfied identically for the plane
trajectories $\dot{\phi} = 0$, while the first equation in the
system~\eqref{berry:eul-th} can be reduced to Eq.~\eqref{berry:eul-li}
considered above in the Lorentz-invariant $\sigma$-model.  Consequently, in this
case $\Gamma(\theta, \phi)$ does not effect on the form of the function $\theta
= \theta(\tau)$ in the instanton solution, but changes the imaginary component
of the Euclidean action.  This effect will be considered in more details in the
section~\ref{s:422}.

If, however, $\Gamma(\theta, \phi_k^{(0)}) \neq 0$, the instanton does not
correspond to a plane solution $\phi = \const$ any longer, and we must seek the
general solution of the system~\eqref{berry:eul} in the form $\theta =
\theta(\tau)$, $\phi = \phi(\tau)$.  In this case the functions $\theta(\tau)$
and $\phi(\tau)$ may in general turn out to be complex-valued.  There are no
general analytical methods for constructing such separatrix solutions; an
instanton solution of the system of equations~\eqref{berry:eul} can be written
exactly only for some cases (see Ref.\cite{IvKir99fnt} and the
section~\ref{s:422} in the present paper).

It will be shown below that the effect of the term in the Lagrangian on the
imaginary component of the Euclidean action $\mathcal{I}$ may lead to nontrivial
consequences even for antiferromagnets with $\Gamma(\theta, \phi) \neq 0$, but
$\Gamma(\theta, \phi_k^{(0)}) = 0$, and there exists a real-valued instanton
solution $\theta = \theta(\tau)$, $\phi = \phi_k^{(0)}$, or in the case when the
value of $H^\text{eff} / H_{ex}$ is negligibly small or its inclusion changes
$\theta = \theta(\tau)$ and the real component of $\mathcal{I}$ insignificantly.

In order to explain this fact, we consider the case when the value of $\Gamma /
H_{ex} \ll 1$ is so small that instanton trajectories can be regarded as planar,
$\theta = \theta(\tau)$, $\phi = \const$.  The presence of the term linear in
$d\bm{l} / d\tau$ leads to the contribution to the imaginary component of the
Euclidean action $\mathcal{I}$, which is proportional to the number of spins in
the particle.  The imaginary component of the Euclidean action $\mathcal{I}$ is
of order of $\im\mathcal{I} / \hbar \propto Nd / J$; i.e., it is
proportional to the product of a small and a large parameter.  Consequently, the
effects of destructive interference can be significant.  It is well known that
the interference effects for orthorhombic ferromagnets may suppress tunneling
completely.\cite{ChudnTej98b, QTM95, LossDiVGrinst92, DelftHenley92} In contrast
to the case of antiferromagnets the term with $d \bm{m} / d \tau$ for
ferromagnets does not contain a small factor $H^\text{eff} / H_{ex}$, but it is
inessential since the value of $\im\mathcal{I} / \hbar \simeq \pi N s \gg 1$ for
ferromagnets, while tunneling is completely suppressed when $\im\mathcal{I} /
\hbar \simeq \pi$.  This condition can easily be satisfied for a large $N$.  In
particular, for the antiferromagnetic particle of the ferritin with $N \simeq
3500$ the tunneling probability in the magnetic field with interference is an
oscillating function of the field, and the suppression of tunneling can be
observed in fields $H \lesssim 100$\,Oe, see Refs.\cite{GolPopkov95, ChiolLoss97,
  ChiolLoss98}, which are much weaker than the characteristic value of the DM
field $H^\text{DM} = 10^3$ -- $10^5$\,Oe.

On the other hand, the contribution to the real component of the Euclidean
action does not contain the large parameter $N$.  This contribution can be
appreciable, see the next section, but in this case the product of other
parameters, namely, the small quantity $d / J \ll 1$ and the large quantity $d /
\widetilde{\beta} \gg 1$, is significant.  Thus, the terms with $d \bm{l} / d
\tau$ may lead to two types of effects: (i) the emergence of nonplanar instanton
trajectories and complex values of components of $\bm{l}$ on these trajectories;
(ii) the interference of instantons even in the case of plane trajectories with
the real Euclidean action $\mathcal{I}$.

The first effect only takes place when the term $\Gamma(\theta, \phi)$ in
Eq.~\eqref{berry:eul} differs from zero.  Such terms are always important for
the description of the domain wall dynamics in antiferromagnets: they may reduce
the limit velocity of the domain wall motion to a considerable extent and may
also lead to an abrupt change in the wall structure upon a continuous variation
of its velocity.\cite{Hom+89, Hom+90} The subsequent analysis of concrete
instanton solutions will show that the role of such terms in the description of
the instanton structure and tunneling is not so important as in the description
of the domain wall dynamics.  On the other hand, if the function $\Gamma(\theta,
\phi)$ differs from zero, but the function $\phi = \phi_k^{(0)}$ for the given
solution $\Gamma(\theta, \phi_k^{(0)}) = 0$, the domain wall dynamics is trivial
and can be described by Lorentz-invariant expressions.  In this case, the
instanton structure $\theta = \theta(\tau)$ is the same as in the
Lorentz-invariant theory.  However, the situation with instantons is different:
not all features can be described by the function $\theta(\tau)$ and the real
component of the Euclidean action $\mathcal{I}$ only.  It will be shown bellow
that the main contribution from the term $\bm{H}^\text{eff} \cdot [(d \bm{l} / d
\tau) \times \bm{l}]$ is associated precisely with interference processes and is
manifested most clearly exactly when the instanton trajectory is planar; i.e.,
$\Gamma(\theta, \phi^{(0)}) = 0$.

In the case of real-valued trajectories, it is convenient to use the following
approach for calculating the imaginary component of the Euclidean
action.\cite{Iv99} We introduce the vector $\bm{r} = r \bm{l}$ which is not
subjected to the condition $\bm{r}^2 = 1$ and present the term with the first
derivative in the expression~\eqref{berry:lagr} in the form
\begin{equation}\label{berry:A}
-i \gamma {\bm{\mathcal{A}}} \frac{\partial\bm{r}}{\partial\tau}
\;, \quad\text{ with } \quad 
\bm{\mathcal{A}} = 
\frac{\bm{r} \times \bm{H}^\text{eff}}{r^2} \;.
\end{equation}
This expression has the same structure as the term in the nonrelativistic
Lagrangian describing the interaction of a classical charged particle moving in
a $3D$ space with coordinate $\bm{r}$ and velocity $\bm{v} = d \bm{r} / d \tau$
with a formal magnetic field $\bm{B} = \bm{\nabla}\times \bm{\mathcal{A}}$
(differentiation is carried out in the $\bm{r}$ space).  It is well known that
the magnetic field appears in the Lagrangian of a charged particle through the
vector potential $\bm{\mathcal{A}}$ at point $\bm{r}$, which is defined only to
within a certain gauge, while the field $\bm{B}$ is gauge-invariant.

Simple but cumbersome calculations proved that, for any ferromagnet, this formal
magnetic field $\bm{B}$ may be radial and can be presented in the form
\begin{equation}\label{berry:B}
\bm{B} = 
\frac{\bm{r}}{r^2} B(\theta, \phi) \;, 
\end{equation}
where
\begin{equation}
B(\theta, \phi) = 
2(\bm{H}^\text{eff} \bm{l}) - \frac{\partial H_i^\text{eff}} {\partial l_i} +
\frac{\partial H_i^\text{eff}} {\partial l_k} l_i l_k \;.
\end{equation}
In the absence of the DM interaction the value of $B(\theta, \phi)$ is
determined only by the external field $\bm{H}^{(0)}$, $B(\theta, \phi) =
2(\bm{H}^{(0)}\bm{l})$.  At zero field the value of $B(\theta, \phi)$ is
determined by the DM field $H_i^\text{DM} = D_{i j}(\bm{l})l_j$ and can be
presented in terms of the tensor $D_{i j}$ as
\begin{equation}\label{berry:B'}
B(\theta, \phi) = 3D_{i j} l_i l_j - D_{i i} + D_{i j, k} l_i l_j l_k -
D_{i j, i} l_j \;.
\end{equation}
Here, the comma in the subscript in $D$ indicates the differentiation of the
tensor $D_{i j}$ with respect to the corresponding component of $\bm{l}$, and
the summation is extended over double indices.  The values of $B(\theta, \phi)$
for various types of DM interaction and of the configurations of axes are given
in column 7 of the Table~\ref{t:dz-an}.

Phases $\Phi_{k, \bar{k}'}$ defined in Eq.~\eqref{berry:comb} can be presented
in terms of the integrals $\int \! \bm{\mathcal{A}} \, d \bm{r}$ taken over
instanton-antiinstanton pairs forming closed loops.  Using the Stokes theorem,
we can present the phase difference $\Phi_{k, k'}$ as the magnetic flux of the
vector $\bm{B}$ through a part of the unit sphere bounded by such a loop.
Obviously, individual phases are determined by the vector potential
$\bm{\mathcal{A}}$, i.e., depend on the gauge, but the phase differences are
gauge-invariant for all pairs of trajectories.

It is important that the structure of $B(\theta, \phi)$ for all possible types
of DM interaction is such that the total flux of field $\bm{B}$ through the unit
sphere,
\begin{equation}
\Phi_{\rm tot} = 
\int_0^\pi \sin\theta d\theta \int_0^{2\pi} d\phi\, B(\theta, \phi)
\end{equation}
is equal to zero and the value of $\cos\Phi_{k, \bar{k}'}$ is independent of the
total time derivative in the Lagrangian (gauge invariance).\footnote{It should
  be noted that the situation in this case is basically different from the case
  of a particle with an uncompensated total spin $S$, where we have the
  potential of the field of the magnetic monopole for $\bm{\mathcal{A}}$ (see
  Eq.~\eqref{lagr:l-v} above).  This vector potential $\bm{\mathcal{A}}$ can be
  written only in a singular form with a singularity on the half-line emerging
  from the point of location of the monopole (the Dirac string).  The total flux
  $\Phi_{tot}$ through the sphere is equal to $4\pi S$, and, hence, the phase
  difference $\Delta\Phi$ for two diametrically positioned trajectories is equal
  to $2\pi S$.  For this reason, the phase factor $\cos(\Delta \Phi / 2) = 0$
  for half-integral values of $S$, and tunneling is forbidden.  It should also
  be noted that, having repeated Dirac's analysis concerning the uniqueness of
  the electron wave function is the monopole field, we can derive the condition
  $\cos(\Phi_{\rm tot} / 2) = 0$ leading to the half-integer quantization of
  uncompensated spin (an example of the quantization of parameters; see
  Ref.\cite{Jackiw84}).  In our case of an uncompensated antiferromagnetic
  particle in the presence of an external field and/or DM interaction,
  $\Phi_{\rm tot} = 0$ and, naturally, no quantization of parameters takes
  place.}

This approach enables us to calculate specifically the phase difference of
integral trajectories and the combinatorial factor $K$ in
Eq.~\eqref{berry:split}.  We begin with the simplest case of an orthorhombic
antiferromagnet for which there are only two pairs of instanton trajectories.
It can be easily verified that the DM interaction of any type (see the
Table~\ref{t:dz-an}) makes a zero contribution to the phase difference for two
diametrically opposite trajectories.\cite{Iv99} For this reason the required
phase factor can be determined only by the external field and can be written in
the form
\begin{equation}\label{berry:cos-field}
\cos\Phi = 
\cos\biggr( \frac{ g \mu_B H^{(0)} N s} {J z} \cos\alpha \biggl) \;,
\end{equation}
where $\alpha$ is the angle between the plane containing the instanton
trajectories and the external field $\bm{H}^{(0)}$.  This result was obtained by
Chiolero and Loss\cite{ChiolLoss98} in particular cases $\alpha = 0$ and $\alpha
= \pi / 2$.  Thus, for instanton trajectories lying in the same plane all
possible types of DM interaction given in the Table~\ref{t:dz-an} do not effect
on the tunneling.  This results does not contradict the analysis carried out by
Golyshev and Popkov\cite{GolPopkov95}, who studied tunneling in small completely
compensated particles of an orthorhombic antiferromagnet with the orthoferrite
structure and discovered that no interference effects exist in a zero magnetic
field.  The our approach enabled us to obtain this result without solution of
the Euler-Lagrange equations.  Thus, we have proved that the conclusion that no
interference takes place for diametrically opposite trajectories can be extended
to more general cases of the DM interaction.  It is important that this
conclusion is drawn without resorting to any approximation, which is inevitable
in the solution of a complex system of equations describing the instanton
structure.

Thus, in the case of orthorhombic antiferromagnets with two instanton
trajectories none of the types of DM interaction presented in the
Table~\ref{t:dz-an} leads to destructive interference.  However, this result is
different for uniaxial antiferromagnets with an easy magnetization axis $C_n$,
$n > 2$.  In this case there exist $n$ pairs of instanton trajectories.
Obviously, here we also have pairs of trajectories lying in the same plane, for
which $\Phi_{k, \bar{k}'} = 0$ and interference is trivial, but it can be
manifested itself for pairs of trajectories with $\phi_k^{(0)} - \phi_{k'}^{(0)}
\neq \pi$.

It will be shown below that the value of combinatorial factor can be reduced
considerably from its maximum value $n^2$ to zero; i.e., both partial and
complete suppression of interference are possible.

\subsection{Tetragonal antiferromagnets}

Let us demonstrate it using specific examples of particles with a tetragonal
easy magnetization axis and binary axes in the perpendicular plane
(crystallographic class $4_z 2_x 2_{x y}$), when the minimum of the real
component of the Euclidean action corresponds to four instanton and four
antiinstanton trajectories.  In order to describe tetragonal antiferromagnets,
we choose the polar axis $\hat{\bm{e}}_3$ along the tetragonal easy
magnetization axis $4_z$.  Anisotropy in the basal plane is determined by the
fourth-order invariant, see the Table~\ref{t:dz-an}.  We assume that $\beta_4 >
0$; i.e., instanton trajectories correspond to the rotation of $\bm{l}$ in the
equivalent planes $z x$ and $z y$.  Depending on the magnetic parity of the
principal axis (according to Turov\cite{Turov65b}) and of the binary axes $2_x$,
$2_y$, or $2_{x y}$, $2_{y x}$ basically different types of behavior can be
observed.  We will consider them separately.

\subsubsection{System $4_z^{(+)} 2_x^{(-)} 2_{xy}^{(-)}$}

With such a structure of axes in an antiferromagnet only the antisymmetric
invariant $d(m_x l_y - m_y l_x)$ is usually considered, which can be obtained
from the antisymmetric component of the tensor of exchange constants.  The value
of $d$ is of order of $\sqrt{\beta J}$, see Ref.\cite{Moriya60}.  This invariant
determines the weak isotropic ferromagnetic moment when $\bm{l}$ is oriented in
the basal plane.  However, it is of no interest to us since it can be reduced to
the total derivative in the Lagrangian and, hence, gives $\Gamma(\theta, \phi) =
0$ in the equations of motion and $B(\theta, \phi) = 0$ in the imaginary part of
the Euclidean action.  In addition, there exist a number of invariants of the
relativistic origin,\cite{Dzyal57} which give a nonzero value of $\Gamma(\theta,
\phi)$.  The simplest of these invariants has the form $2 (l_x^2 - l_y^2) (m_x
l_y + m_y l_x)$ which coincides (except for the total derivative) with the
invariant $(m_+ l_+^3 - m_- l_-^3) / (2i)$ presented in the table.  It can
easily be seen, however, that in this case $\Gamma(\theta, \phi_k^{(0)}) = 0$,
and the instanton solution has the form $\theta = \theta(\tau)$, $\phi =
\phi_k^{(0)} = \pi k / 2$ for integer $k$.  The value of $B(\theta, \phi)$ is
such that
\begin{equation}\label{berry:B-4+}
B \propto 
\sin^4 \theta\sin4\phi \;,
\end{equation}
and, hence, all phases $\Phi_{k, \bar{k}'}$ are equal to zero.  An analysis of
other invariants, e.g., $l_z^2 (m_x l_y - m_y l_x)$, leads to the same result
(namely, the DM interaction does not effect on tunneling in any way).  This
result is apparently independent of the model and is determined only by the type
of magnetic symmetry.  The model independence for dynamic effects in domain
walls has the same origin, i.e., the DM interaction; it was demonstrated in
Refs.\cite{Hom+89, Hom+90}.  Thus, the case of an even principal axis
$4_z^{(+)}$ may serve as an example that nonzero terms which are linear in $d
\bm{l} / d \tau$ and cannot be reduced to the total derivative are not appeared
in the separatrix solution and do not effect on tunneling in any way.

A different situation takes place for antiferromagnets with an odd principal
axis $4_z^{(-)}$. In this case two cases are possible: when the intermediate
anisotropy axes through which tunneling takes place are odd and when these axes
are even.

\subsubsection{System $4_z^{(-)} 2_x^{(-)} 2_{xy}^{(+)}$}

In this case $\Gamma(\theta, \phi) = 0$ and $\phi = \phi_k^{(0)} = \pi k / 2$.
The presence of the DM interaction does not effect on the instanton trajectories
with the minimum action, which correspond to $\phi = \phi_k^{(0)}$, $\theta =
\theta(\tau)$.  However, in contrast to the system with $4_z^{(+)}$, the
contribution of the DM interaction is significant for calculating the
combinatorial factor $K$ in Eq.~\eqref{berry:split}.  It can be seen from the
explicit expression $B(\theta, \phi) \propto \sin^2\theta \sin2\phi$ that the
phase difference for adjacent trajectories (with $\phi = \phi_k^{(0)}$ $\phi =
\phi_{k \pm 1}^{(0)}$) differs from zero.

Thus, the phase factor for the tunneling probability is given by
\begin{equation}
K = 
16 \sin^2 \biggl( \frac{s d N}{J} \biggl) \;.
\end{equation}
It is an oscillating function of the DM interaction constant $d$, and it takes
the values in the range from $0$ to $16$.  For realistic values of $N$ of order
of $10^3$ -- $10^5$ the period is not large; the value of $\Delta H^\text{DM} /
H^\text{DM} \simeq 10^{-3} \div 10^{-1}$ for characteristic values of
$H^\text{DM} \simeq 10^4$\,Oe and $H_{ex} \simeq 10^6$\,Oe.  Since the value of
the DM field is very sensitive to extrinsic parameters (e.g., the pressure or
the addition of a small amount of impurities to the crystal), these oscillations
can be observed and monitored.  An additional opportunity for observing
interference effects appears when the magnetic field is taken into
consideration.

\subsubsection{System $4_z^{(-)} 2_x^{(+)} 2_{xy}^{(-)}$}

Such a symmetry group is typical for the extensively studied weak
antiferromagnet MnF$_2$, see Ref.\cite{ShapZak68}.  In this case $\Gamma(\theta,
\phi) \propto \sin^3\theta \cos2\phi$ and $\Gamma(\theta, \phi) \neq 0$ for all
values of $\phi = \phi_k^{(0)}$, corresponding to the minimum of the anisotropy
in the basal plane and describing instanton trajectories for $d = 0$.  For $d
\ne 0$ instanton solutions cannot be written in the simple form $\theta =
\theta(\tau)$, $\phi = \pi k / 2$, $k = 0, 1, 2, 3$.  On the other hand, if we
assume that the value of $d$ is very small, we can easily find, applying the
approximation of planar rotation and using the formula $B(\theta, \phi) \propto
\sin^2\theta \cos2\phi$, that the difference in the imaginary components of
$\mathcal{I}$ for pairs of trajectories lying in the same plane as well as for
adjacent instanton trajectories is equal to zero, and no interference effects
take place.  We will consider the solution for this case in the next section and
prove that these simple regularities are preserved in a more rigorous analysis
also when we do not require that $\phi = \pi k / 2$.  We will also consider a
general mechanisms of the tunneling in the case when $\Gamma(\theta,
\phi^{(0)}_{\text{min}}) \ne 0$ and the instanton solution is not real-valued.

\subsection{Hexagonal antiferromagnets}

Let us briefly consider the case of a hexagonal principal axis.  Here, we again
have three cases presented in the Table~\ref{t:dz-an}.  For a system with an
even principal axis $6_z^{(+)} 2_x^{(-)} 2_{\pi / 6}^{(-)}$ there exists the
invariant $m_x l_y - m_y l_x$ and the DM interaction, which cannot be reduced to
a total derivative, appears only in the fifth order in $\bm{l}$.  The analysis
is similar to that for the system $4_z^{(+)} 2_x^{(-)} 2_{xy}^{(-)}$.  In this
case also nonzero terms which are linear in $d \bm{l} / d \tau$ and cannot be
reduced to a total derivative do not effect on tunneling in any way.  It can be
verified that such a behavior is the same as for antiferromagnets with the even
principal axis $n_z^{(+)}$.

For systems with an odd principal axis, the DM interaction is a cubic function
of $\bm{l}$, but it makes a zero contribution to the imaginary component of the
Euclidean action under the assumption that instanton trajectories are real and
planar.  In the system $6_z^{(-)} 2_x^{(-)} 2_{\pi / 6}^{(+)}$ minimal instanton
trajectories have an imaginary component, but this only changes the real part of
the Euclidean action.  The analysis of this system is similar to that which will
be carried out in the section~\ref{s:422} for the system $4_z^{(-)} 2_x^{(+)}
2_{xy}^{(-)}$.  Consequently, the combinatorial factor $K$ for all three cases
has the maximum value equal to $K = 36$.


\section{Instanton solution for antiferromagnets with the symmetry 
  $4_z^{(-)} 2_x^{(+)} 2_{xy}^{(-)}$} \label{s:422}

It was noted above that in the case of antiferromagnets with the symmetry
$4_z^{(-)} 2_x^{(+)} 2_{xy}^{(-)}$ there is no exact solution of the type $\phi
= \pi k / 2$, $\theta = \theta(\tau)$ for trajectories with the rotation of
$\bm{l}$ in the vicinity of the easy plane $\phi \simeq \pi k / 2$, and we have
to analyze the complete system of two equations~\eqref{berry:eul-th} and
\eqref{berry:eul-phi}.  The situation is complicated further since these
equations have complex-valued coefficients and, in general, their complex
solutions of the type $\theta = \theta_1(\tau) + i \theta_2(\tau)$, $\phi =
\phi_1(\tau) + i \phi_2(\tau)$ must be considered.  As a result,
system~\eqref{berry:eul} is equivalent to a dynamic system with four degrees of
freedom, and it is not integrable.  No general method exists for an analysis of
such systems.  However, a comprehensive analysis can be carried out in the given
case as well as for magnets with other types of symmetry, which are listed in
the Table~\ref{t:dz-an}.

In the case of antiferromagnets with the odd tetragonal axis, the equations for
the angular variables $\theta$ and $\phi$ have the form
\begin{subequations}\label{422:eul}
\begin{multline}\label{422:eul-th}
-\ddot{\theta} + \dot{\phi}^2 \sin\theta \cos\theta +
\omega_0^2 \sin\theta \cos\theta(1 + \beta_4 \sin^2\theta \sin^2 2\phi) \\
+3 i \omega_D \dot{\phi} \sin^3\theta \cos2\phi = 0 \;,
\end{multline}
\begin{multline}\label{422:eul-phi}
-\ddot{\phi} \sin^2\theta - 2\dot{\phi} \dot{\theta} \sin\theta \cos\theta +
\beta_4 \omega_0^2 \sin^4\theta \sin2\phi \cos2\phi \\
-3 i \omega_D \dot{\theta} \sin^3\theta \cos2\phi = 0 \;.
\end{multline}
\end{subequations}

The quantity $\omega_0$ defines the height of the potential barrier through
which tunneling occurs, and $\beta_4$ is the dimensionless anisotropy constant
in the basal plane.  We assume that $\beta_4 > 0$, which corresponds to
instantons of the Lorentz-invariant $\sigma$-model ($\omega_D = 0$) that passing
through the even axes $x$ or $y$; $\beta_4 \ll 1$ corresponds to the easy-axis
limit, and $\omega_D$ is proportional to the DM interaction constant, $\omega_D
= \gamma |\bm{H}^{\rm DM}| = z d / \hbar$.

It can be easily seen that this system has the exact solution $\phi = \pi (2k +
1) / 4$, $\theta = \theta(\tau)$, which has been considered in the previous
section.  It determines tunneling for $\beta_4 < 0$, but in the case of $\beta_4
> 0$ we are interested in now, it corresponds to the rotation of $\bm{l}$ in the
hard planes, does not ensure that the real part of the Euclidean action has a
minimum, and makes a zero contribution to the tunneling amplitude in the
instanton approximation.  The exact solution $\phi = \pi k / 2$ does not exist
in this case.  It can be seen, however, that the substitution $\phi = \pi k / 2
+ i f(\tau)$ and $\theta = \theta(\tau)$ with the real functions $f(\tau)$,
$\theta(\tau)$ and does not contradict this system and leads to the following
system of two equations for functions $f$ and $\theta$:
\begin{subequations}\label{422:eul1}
\begin{multline}\label{422:eul1-th}
-\ddot{\theta} - \dot{f}^2 \sin\theta \cos\theta +
\omega_0^2 \sin\theta \cos\theta (1 - \beta_4 \sin^2\theta \sinh^22f) \\
-3 (-1)^k \omega_D \dot{f} \sin^3 \theta\cosh2f = 0 \;,
\end{multline}
\begin{multline}\label{422:eul1-phi}
\ddot{f} \sin^2\theta + 2\dot{f} \dot{\theta} \sin\theta \cos\theta-
\beta_4 \omega_0^2 \sin^4\theta \sinh2f \cosh2f \\
+3 (-1)^k \omega_D \dot{\theta} \sin^3\theta \cosh2f = 0 \;.
\end{multline}
\end{subequations}

In addition, such a substitution renders the Lagrangian real-valued:
\begin{multline}\label{422:lagr-re}
\mathcal{L} = 
\frac{\hbar^2N}{2 J z}\biggl[
\frac{\dot{\theta}^2 - \dot{f}^2 \sin^2\theta}{2} + \\
(-1)^k \omega_D \dot{f} \cosh2f \cos\theta (2 + \sin^2\theta) \\
+\frac{\omega_0^2}{2} \sin^2\theta -
\frac{\omega_0^2 \beta_4}{4} \sin^4\theta \sinh^2 2f \biggr] \;.
\end{multline}

It can be seen that the imaginary component $f = \im\phi$ of the instanton
solution effects only on the real part of the Euclidean action.  The
system~\eqref{422:eul1} is equivalent to a mechanical system with two degrees of
freedom.  Only one first integral is known for it:
\begin{multline}\label{422:1int}
\mathcal{E} = 
\frac{\hbar^2N}{2 J z}\biggl[
\frac{\dot{\theta}^2 - \dot{f}^2 \sin^2\theta}{2}-
\frac{\omega_0^2}{2} \sin^2\theta \\
+ \frac{\omega_0^2\beta_4}{4} \sin^4\theta \sinh^22f \biggr] \;.
\end{multline}
Note that $\mathcal{E} = 0$ for the separatrix solutions we are interested in.
For this reason this system cannot be analyzed exactly.  However, an approximate
solution can be constructed in the physically interesting case, when $\omega_D
\ll \omega_0$, $\beta_4 \ll 1$ and for any relation between $\omega_D$ and
$\omega_0\beta_4$.  In order to find such a solution, we note that
Eq.~\eqref{422:eul1-th} from the system~\eqref{422:eul1} is transformed into
$(\dot{\theta})^2 = \omega_0^2 \sin^2\theta$ in the zeroth approximation in the
small parameters $\omega_D / \omega_0$ and $\beta_4$. In this case the constant
solution $f = f_0 = \const$ satisfies Eq.~\eqref{422:eul1-phi} and gives
\begin{equation}\label{422:0app}
\sinh 2 f_0 = 
\frac{\omega_D}{\beta_4 \omega_0} \;.
\end{equation}
It should be noted that the value of $f_0$ is determined by the ratio of two
small parameters and can be appreciable.  Using this fact, we can write a
refined equation for $\theta(\tau)$
\begin{equation}\label{422:1app}
\ddot{\theta} = 
\omega_0^2 \sin\theta \cos\theta \biggl(1 - 
\frac{\omega_D^2}{\beta_4 \omega_0^2} \sin^2\theta \biggr) \;.
\end{equation}

The approximate solution constructed by us is valid if $\dot{\theta} \simeq
\omega_0 \sin\theta$, i.e.
\begin{equation}\label{422:om-d}
\omega_D^2 \ll 
\frac{\omega_0^2}{\beta_4} \;.
\end{equation}

This condition may also hold in the case when the value of $\sinh 2f_0 =
\omega_D / \beta_4 \omega_0$ is of order of unity, but it is still violated for
$\beta_4 \to 0$.  In this case the situation is similar to that observed for the
domain wall dynamics:\cite{Hom+90} the limit velocity of a domain wall in a
tetragonal antiferromagnet with an odd axis with $\phi \neq \const$ tends to
zero as $\beta_4 \to 0$, and no dynamic solution exists for $\beta_4 = 0$.

The value of the Euclidean action for the obtained solution is real and is
defined by the expression
\begin{equation}\label{422:act}
\mathcal{I} = 
\frac{\hbar^2 N \omega_0}{J z}
\biggl(1 + \frac{\omega_D^2}{3 \beta_4 \omega_0^2} \biggl) \;.
\end{equation}
In the range of applicability of the constructed solution, i.e., when $\omega_D$
and $\beta_4$, are small and when inequality~\eqref{422:om-d} is satisfied, the
inclusion of the DM interaction leads only to a small correction to the real
part of the Euclidean action, the imaginary part being identically equal to
zero.

For models of antiferromagnets with binary and hexagonal symmetry axes such
an approximate solution cannot be constructed, but the analysis of these models
is even simpler than in the case of antiferromagnets with a tetragonal
symmetry axis.  In these cases we can also verify that, if the exact solution
$\phi = \phi_k^{(0)} = 2 \pi k / n$, $\theta = \theta(\tau)$ does not exist, the
solution has the same form as before:
\begin{equation}\label{422:1app-gen}
\phi = 
\frac{2 \pi k}{n}+ i f(\tau) \;, \qquad 
\theta = 
\theta(\tau) \;,
\end{equation}
and the term in the Lagrangian which is linear in $d \bm{l} / d \tau$
contributes only to the real part of Euclidean action $\mathcal{I}$.  It can be
proved, however, that function $f(\tau)$ is antisymmetric and proportional to
the parameter $\omega_D / (\beta \omega_0)$, which is always small (in contrast
to the case of a tetragonal antiferromagnet, in which there appears the
parameter $\omega_D / (\beta_4 \omega_0)$ whose value may be large).
Consequently, we can seek the function $f \ll 1$ using the same perturbation
theory as for the domain wall dynamics in such antiferromagnets.\cite{Hom+90} As
a result, we obtain the following expression for the real part of the Euclidean
action:
\begin{equation}\label{422:2,6}
\mathcal{I} = 
\frac{\hbar^2 N \omega_0}{J z} \biggl (1 + \xi
\frac{\omega_D^2}{\omega_0^2} \biggl) \;,
\end{equation}
where the numerical factor $\xi$ is of order of unity.  Thus, the correction to
the result typical for the Lorentz-invariant model is always small.  Note that
no interference effects take place in this case.  We arrive at the conclusion
that the DM interaction effects on the tunnel probability in hexagonal and
orthorhombic antiferromagnets.

\section{Concluding remarks} \label{s:conclusion}

The analysis of antiferromagnetic particles with a tetragonal symmetry axis
shows that three possible types of the effect of the DM interaction on tunneling
processes exist.  The investigation of the remaining cases important for the
analysis of crystalline antiferromagnets (orthorhombic or uniaxial with a
hexagonal symmetry axis, see the Table~\ref{t:dz-an}) proved that these types
include all possible cases for antiferromagnetic systems with a doubly
degenerate ground state.  In fact, all cases can be reduced to the following
three types of behavior.

\begin{enumerate}
  
\item The principal axis is even, e.g., $4_z^{(+)}$ or $6_z^{(+)}$. In this case
  vector $\bm{l}$ is real on all instanton trajectories and these trajectories
  are planar $\phi = 2 \pi k / n$, $\theta = \theta(\tau)$.  The real part of
  the Euclidean action is independent of the DM interaction constant, the
  imaginary part is equal to zero, and destructive interference effects are
  absent.  In this case tunneling can be described without taking into account
  the DM interaction.
  
\item The principal axis is odd, and there exists an exact real solution with
  the rotation of $\bm{l}$ in the easy plane determined by the anisotropy.  Such
  an example is a system of the type $4_z^{(-)} 2_x^{(-)}$, in which the
  instanton trajectory is plane and the DM constant does not appear in the real
  part of the Euclidean action.  In this case, however, the inclusion of the DM
  interaction leads to the emergence of the imaginary part of the Euclidean
  action and may effect on the tunneling probability due to the interference of
  instanton trajectories lying in different planes.  Since the corresponding
  phase factor contains the large value $N$, tunneling can be suppressed
  completely due to the destructive interference even for small values of the DM
  constant.

\item The principal axis is odd, and the simple solution $\phi = 2 \pi k / n$
  does not exist.  In this case, the vector $\bm{l}$ has both real and imaginary
  components, but all types of DM interaction change only the real part of the
  Euclidean action, this change being small in view of the smallness of the
  parameter $d^2 / (J \beta)$.  The imaginary part of the Euclidean action is
  equal to zero and destructive interference effects are absent.

\end{enumerate}

Thus, the only important effect produced by the DM interaction is associated
with the possibility of the interference of instanton trajectories in the case
when their number is greater than two (an antiferromagnetic particle with the
easy magnetization axis approximately corresponds to $n > 2$).  This effect can
be observed for antiferromagnets with an odd principal axis in the case when the
rotation of $\bm{l}$ on the instanton trajectory also occurs through the odd
axis.  It is associated with the interference of pairs of instanton trajectories
lying in different planes.  Since $\bm{l}$ is real in this case and all
instanton trajectories are plane, an exact analysis can easily be carried out
and the description of tunneling is reduced to the geometrical analysis
described in the section~\ref{s:berry}.

\begin{acknowledgments}
  The authors thank A.K.~Kolezhuk for fruitful discussions.  This work was
  supported financially by Volkswagen Stiftung (grant No.~I/75895).
\end{acknowledgments}

\end{document}